\documentclass[a4paper,12pt]{article}
\pdfoutput=1
\usepackage{test}

\usepackage{xcolor}
\definecolor{pantone}{RGB}{1, 33, 105}

\usepackage{graphicx}
\usepackage{hyperref}
\hypersetup{colorlinks,linkcolor=blue,citecolor=blue,urlcolor=pantone,hypertexnames=false}

\usepackage{booktabs,caption}
\usepackage{longtable}
\usepackage{array}

\newcolumntype{L}[1]{>{\raggedright\arraybackslash}p{#1}}

\usepackage[style=ext-authoryear-comp,
sorting=nyt,
dashed=false,
maxcitenames=2,
maxbibnames=99,
uniquelist=false,
uniquename=false,
giveninits=true,
natbib,
date=year
]{biblatex}

\AtBeginRefsection{\GenRefcontextData{sorting=ynt}}
\AtEveryCite{\localrefcontext[sorting=ynt]}

\DeclareFieldFormat{pages}{#1}
\renewbibmacro{in:}{\ifentrytype{article}{}{\printtext{\bibstring{in}\intitlepunct}}}
\DeclareFieldFormat[article,inbook,incollection,inproceedings,patent,thesis,unpublished]{titlecase:title}{\MakeSentenceCase*{#1}}

\newbibmacro*{atoda:titlelink}[1]{%
  \ifhyperref
    {\iffieldundef{doi}
       {\iffieldundef{eprint}
          {\iffieldundef{url}
             {#1}
             {\href{\thefield{url}}{#1}}}
          {\href{https://arxiv.org/abs/\thefield{eprint}}{#1}}}
       {\href{https://doi.org/\thefield{doi}}{#1}}}
    {#1}}

\DeclareFieldFormat{title}{\usebibmacro{atoda:titlelink}{\mkbibemph{#1}}}
\DeclareFieldFormat
  [article,inbook,incollection,inproceedings,patent,thesis,unpublished]
  {title}{\usebibmacro{atoda:titlelink}{\mkbibquote{#1\isdot}}}
\renewbibmacro*{doi+eprint+url}{}

\usepackage[inline,shortlabels]{enumitem}
\setlist[enumerate,1]{label=(\roman*)}

\usepackage[margin=1.25in]{geometry}
\usepackage{setspace}
\renewcommand{\Pr}{\operatorname{P}}

\newcommand{\imp}{\mathrm{imp}}
\newcommand{\code}{\mathrm{code}}

\numberwithin{equation}{section}

\newcommand{\numLadderGiniTenK}{0.691}
\newcommand{\numLadderMedianTenK}{0.4055}
\newcommand{\numLadderPninenineTenK}{8.46}
\newcommand{\numLadderTailTenK}{1.383}
\newcommand{\numLadderGiniHundredK}{0.750}
\newcommand{\numLadderMedianHundredK}{0.3315}
\newcommand{\numLadderPninenineHundredK}{7.92}
\newcommand{\numLadderTailHundredK}{1.350}
\newcommand{\numLadderGiniMillion}{0.697}
\newcommand{\numLadderMedianMillion}{0.3937}
\newcommand{\numLadderPninenineMillion}{8.92}
\newcommand{\numLadderTailMillion}{1.400}
\newcommand{\numHorizonXsGiniMeanRange}{0.052}
\newcommand{\numHorizonDynGiniDrop}{0.19}

\newcommand{\numHorizonLongDynTail}{1.510}
\newcommand{\numHorizonLongDynTailMean}{1.474}

\newcommand{\numSigmaBarRatio}{43}
\newcommand{\numXsMean}{1.0002}
\newcommand{\numPooledMean}{1.0001}
\newcommand{\numLadderMeanTenK}{1.0001}
\newcommand{\numLadderMeanHundredK}{1.0011}
\newcommand{\numLadderMeanMillion}{1.0004}
\newcommand{\numXsLadderGiniTenK}{0.892}

\newcommand{\numXsLadderMeanHundredK}{---}
\newcommand{\numXsLadderMedianHundredK}{0.0312}
\newcommand{\numXsLadderPninenineHundredK}{---}
\newcommand{\numXsLadderGiniHundredK}{0.860}
\newcommand{\numXsLadderTailHundredK}{---}
\newcommand{\numXsLadderMeanMillion}{---}
\newcommand{\numXsLadderMedianMillion}{0.0052}
\newcommand{\numXsLadderPninenineMillion}{---}
\newcommand{\numXsLadderGiniMillion}{0.883}
\newcommand{\numXsLadderTailMillion}{---}
\newcommand{\numArgumentFfour}{0.187}
\newcommand{\numXsGini}{0.892}

\newcommand{\numXsMedian}{0.0304}
\newcommand{\numXsPninenine}{10.43}
\newcommand{\numXsTail}{1.060}
\newcommand{\numXsHill}{0.922}
\newcommand{\numPooledGini}{0.871}
\newcommand{\numPooledMedian}{0.0241}
\newcommand{\numPooledPninenine}{11.06}
\newcommand{\numPooledTail}{1.204}
\newcommand{\numDynGini}{0.691}

\newcommand{\numBothGini}{0.914}
\newcommand{\numBothMedian}{0.0198}

\newcommand{\numBothMean}{1.0000}
\newcommand{\numBFMean}{1.0002}
\newcommand{\numCorrGiniMean}{0.879}
\newcommand{\numCorrGiniSD}{0.029}
\newcommand{\numCorrGiniMin}{0.837}
\newcommand{\numCorrGiniMax}{0.946}
\newcommand{\numCorrTailMean}{1.201}
\newcommand{\numCorrTailSD}{0.109}
\newcommand{\numCorrHillMean}{1.184}
\newcommand{\numCorrHillSD}{0.182}

\newcommand{\numDynGiniSD}{0.038}
\newcommand{\numDynGiniMin}{0.655}
\newcommand{\numDynGiniMax}{0.802}
\newcommand{\numDynTailMin}{1.292}
\newcommand{\numDynTailMax}{1.466}
\newcommand{\numHighFactorCount}{6.4}
\newcommand{\numAbsPayoffCount}{0.4}

\newcommand{\numKgridTwentyfive}{0.684}
\newcommand{\numHorizonSeeds}{16}
\newcommand{\numHorizonShortYears}{50}
\newcommand{\numHorizonShortDynGini}{0.758}

\newcommand{\numHorizonBaseDynGini}{0.691}

\newcommand{\numHorizonLongYears}{1{,}000}
\newcommand{\numHorizonLongDynGini}{0.570}

\newcommand{\numHorizonXsGiniRange}{0.129}
\newcommand{\numHorizonDynGiniSDmax}{0.039}
\newcommand{\numHorizonXsGiniSDmax}{0.040}
\newcommand{\numHorizonLongDynGiniMean}{0.587}
\newcommand{\numHorizonLongDynGiniSD}{0.032}
\newcommand{\numSigmaBarSq}{0.0165}
\newcommand{\numMeanEta}{0.0009}
\newcommand{\numDeltaOverSigmaSq}{2.3\times 10^{-2}}
\newcommand{\numArgumentFfive}{3.7\times 10^{-3}}
\newcommand{\numMuFfour}{1.0447}
\newcommand{\numMuFfive}{1.0009}
\newcommand{\numSigmaBarSqNeededFfour}{1.4\times 10^{-3}}
\newcommand{\numSigmaBarSqNeededFfive}{2.7\times 10^{-5}}
\newcommand{\numSigmaBarSqBF}{0.0166}
\newcommand{\numMuFfourBF}{1.0442}
\newcommand{\numMuFfiveBF}{1.0009}

\newcommand{\numMuBlock}{0.042}
\newcommand{\numInverseLambdaWeeks}{51.0}
\newcommand{\numInverseLambdaYears}{0.98}
\newcommand{\numBeliefSDyoung}{2.0\times 10^{-1}}
\newcommand{\numBeliefSDone}{4.3\times 10^{-2}}
\newcommand{\numBeliefSDfive}{4.1\times 10^{-4}}
\newcommand{\numBeliefSDtwenty}{4.2\times 10^{-11}}
\newcommand{\numSweepSeeds}{16}
\newcommand{\numSigmaBarSqMin}{0.009}
\newcommand{\numSigmaBarSqMax}{3.28}
\newcommand{\numMuFfourSweepMin}{1.000}
\newcommand{\numMuFfourSweepMax}{1.078}
\newcommand{\numTopDateShare}{94}

\InputIfFileExists{comment_numbers}{}{}

\newcommand{\horizonRows}{%
        BF dynasty average, Gini & 0.758 & 0.745 & 0.730 & 0.691 & 0.618 & 0.591 & 0.570 \\
        \quad standard deviation across seeds & 0.025 & 0.032 & 0.033 & 0.038 & 0.033 & 0.039 & 0.032 \\
        BF dynasty average, median$/H$ & 0.242 & 0.314 & 0.350 & 0.405 & 0.493 & 0.512 & 0.533 \\
        \addlinespace
        Terminal cross section, Gini & 0.812 & 0.889 & 0.873 & 0.902 & 0.873 & 0.941 & 0.823 \\
        \quad standard deviation across seeds & 0.024 & 0.029 & 0.035 & 0.031 & 0.035 & 0.040 & 0.026 \\
        Terminal cross section, median$/H$ & 0.0663 & 0.0165 & 0.0247 & 0.0275 & 0.0030 & $<$0.0001 & 0.0621 \\%
}
\newcommand{\correctionRows}{%
        BF program & 1.0002 & 0.0304 & 10.43 & 0.892 & 1.060 & 0.922 \\
        Correct prices only & 1.0000 & 0.0304 & 10.40 & 0.892 & 1.060 & 0.919 \\
        Correct timing only & 1.0002 & 0.0201 & 9.51 & 0.914 & 0.982 & 0.856 \\
        Both corrections & 1.0000 & 0.0198 & 9.76 & 0.914 & 0.982 & 0.876 \\%
}
\InputIfFileExists{comment_table_horizon}{}{}
\InputIfFileExists{comment_table_corrections}{}{}

\title{Self-Fulfilling Prophecies, Quasi Nonergodicity, and Wealth Inequality: A Comment}
\author{Alexis Akira Toda \thanks{Department of Economics, Emory University and Research Institute for Economics and Business Administration, Kobe University. Email: \href{mailto:alexis.akira.toda@emory.edu}{alexis.akira.toda@emory.edu}.}}
\date{}

\begin{document}

\maketitle

\begin{abstract}
\citet{BouchaudFarmer2023} argue that self-fulfilling beliefs generate a realistic wealth distribution. Their reported Gini coefficient, quantiles, and Pareto exponent are computed from dynasty-level wealth averaged over $250$ dates, not from a cross section. Correcting the estimand strengthens their qualitative result: with $N=1{,}000{,}000$ dynasties, the terminal cross-sectional Gini is $\numXsLadderGiniMillion$ rather than $0.7$, and median wealth is $\numXsLadderMedianMillion$ rather than $0.39$ of average wealth. The model therefore generates more inequality, but not the reported quantitative fit. Appendix F analyzes a different process and does not establish the claimed Pareto tail or exponent.
\end{abstract}

\section{Introduction}\label{sec:intro}

\citet{BouchaudFarmer2023} (henceforth BF) study an economy in which agents disagree about a public event whose probability equals the population mean belief. Agents trade a complete set of state-contingent claims. Because prices reflect a wealth-weighted mean belief, each realization redistributes wealth toward agents who assigned that realized outcome relatively high probability. BF report that this mechanism generates realistic wealth inequality: a Gini coefficient of $0.7$, median wealth equal to $39$ percent of average wealth, a $99$th percentile equal to $892$ percent of average wealth, and a Pareto tail with exponent near $1.4$.

This comment shows that BF's wealth-distribution analysis does not support these quantitative conclusions. Correcting the estimand strengthens rather than overturns their central qualitative finding: disagreement and learning generate substantial wealth inequality. Their reported inequality statistics, however, use dynasty-level time averages rather than a cross section, while their Pareto-tail argument analyzes a different stochastic process and fails to reproduce the reported exponent at the model's calibration. Consequently, the empirical comparison rests on a time-averaged estimand and the claimed tail exponent lacks a theoretical foundation in the model. Although the model generates more inequality than BF report, the specific quantitative fit and tail exponent do not follow from their analysis.

The first problem concerns the unit of observation, discussed in Section \ref{sec:dist}. BF's replication program\footnote{See the \href{https://doi.org/10.7910/DVN/R9Y6YN}{BF replication archive}. I examine \texttt{MakeFigures56and7.m}, which BF supply for Figures 5--7. All line numbers refer to the archived version.} records $250$ cross sections but first averages wealth over time within each dynasty before calculating the histogram, quantiles, Lorenz curve, Gini coefficient, and Pareto regression. A dynasty is an index occupied by successive agents as death and replacement occur, not a household observed at one date. At BF's population size, $N=1{,}000{,}000$, the dynasty-average procedure gives a Gini coefficient of $\numLadderGiniMillion$, reproducing their reported $0.70$; the terminal cross section from the same history instead gives $\numXsLadderGiniMillion$. Its median-to-average wealth ratio is approximately $\numXsLadderMedianMillion$: median wealth is about $0.52$ percent of average wealth, rather than the reported $39$ percent.

Time averaging also creates horizon dependence. Under ergodicity with a finite mean, each dynasty's average wealth converges to the common mean, so the distribution of dynasty averages collapses toward equality rather than converging to the stationary cross-sectional distribution. Holding the recorded dates and burn-in fixed, the dynasty-average Gini coefficient falls from $\numHorizonShortDynGini$ at $\numHorizonShortYears$ years to $\numHorizonBaseDynGini$ at $300$ years and $\numHorizonLongDynGini$ at $\numHorizonLongYears$ years; the terminal cross-sectional Gini exhibits no corresponding trend. BF's reported $0.7$ is therefore a feature of the averaging window, not a measure of stationary cross-sectional inequality.

The second problem concerns the upper tail, discussed in Section \ref{sec:tail}. A Pareto tail is a recurring claim in BF. Their Introduction states that the model generates a Pareto-tailed wealth distribution and reproduces the empirical value of its exponent \citep[p.~950]{BouchaudFarmer2023}. Section VIII states that Figure 7 ``reveals a power-law tail'' with exponent $1.4$, reports the same exponent under alternative calibrations, and invokes Appendix F to approximately compute the exponent, establish $\mu>1$ whenever $\delta>0$, and obtain a nondegenerate small-mortality limit \citep[pp.~976, 978--980]{BouchaudFarmer2023}. Appendix F is therefore offered as analytical support for claims made repeatedly in the main text. It instead assumes a multiplicative-reset process whose return is independent of current wealth and renewed after a fixed interval. Neither property holds in the equilibrium model: returns depend on the wealth-weighted belief, all surviving agents respond to the same public signal, and idiosyncratic return differences largely disappear within a few years of birth. Although multiplicative growth combined with death or replacement can generate Pareto tails \citep{DuttaMichel1998,NireiSouma2004,Toda2014JET,NireiAoki2016,BeareToda2022ECMA}, applying those results requires conditions on the stochastic recursion, stationarity, and dependence structure. BF do not establish these conditions, and key assumptions---particularly wealth-independent returns and the treatment of common shocks---are violated in their equilibrium model. At BF's calibration, Equations (F4) and (F5) yield $\mu=\numMuFfour$ and $\mu=\numMuFfive$, respectively, rather than $1.4$, and the persistence correction in (F5) does not follow from its motivating block-constant approximation. Thus, Appendix F establishes neither the claimed Pareto tail nor its exponent for the BF economy.

Appendix \ref{sec:program} documents two additional implementation discrepancies. The pricing calculation combines a finite-population numerator with a large-population denominator, violating the aggregate wealth identity, and beliefs are updated with the realized signal before claims contingent on that signal are settled. At $N=10{,}000$, correcting both changes the cross-sectional Gini coefficient from $\numXsGini$ to $\numBothGini$. These discrepancies therefore do not explain BF's $0.7$; the decisive issue is the distributional estimand.

These corrections leave intact BF's central qualitative mechanism: their self-referential belief process, which combines constant-gain learning with demographic turnover, can sustain disagreement and aggregate belief fluctuations and, when coupled to trading, generate substantial wealth inequality. However, the quantitative characterization of wealth inequality needs to be revised. BF call the Arrow-security allocation rule ``[t]he novel aspect of our approach'' \citep[p.~967]{BouchaudFarmer2023}. \citet[p.~565]{Rubinstein1976} had already derived the same state-contingent log-utility rule and its wealth-weighted consensus belief; \citet{DetempleMurthy1994} and \citet{JouiniNapp2007} provide intertemporal and complete-market formulations. These papers are not cited by BF. Thus the allocation, aggregation, and pricing formulas specialize established results. The distinctive contribution lies in coupling the self-referential belief process to wealth dynamics, not in the allocation formula itself.

\section{BF economy}\label{sec:model}

\subsection{Model summary and relevant sections of BF}\label{subsec:model_summary}

BF's economy has two blocks. The first specifies beliefs. A population of fixed size observes a sequence of binary public events. By assumption, the probability of the next event equals the population mean belief; no external fundamental probability enters the model. Conditional on survival, agents update through constant-gain learning: they put weight $\lambda$ on the latest event and weight $1-\lambda$ on their previous belief, as in BF's Equation~\eqref{eq:BF5}, reproduced below. They die at a constant rate and are replaced by newborns with randomly drawn beliefs. Mortality and constant-gain learning prevent convergence and sustain disagreement.

The second block places these agents in a market. Each agent receives the same perishable income every period and trades a complete set of securities contingent on the public event and mortality outcomes. With logarithmic preferences, agents consume a fixed share of their wealth and invest the remainder in proportion to their beliefs. Two averages of beliefs then coexist. The unweighted average determines the probability of the public event by construction, whereas the wealth-weighted average is reflected in security prices. The two differ whenever wealth is unequally distributed. This difference permits systematic redistribution across agents.

Formally, consider $N$ dynasties indexed by $i$, each containing one agent at every date. Let $p_{i,t}$ denote the belief of the agent in dynasty $i$ that the public signal $s_t\in\set{0,1}$ equals one. BF impose the self-referential restriction in their Equation (1),
\begin{equation}
    P_t\coloneqq \Pr(s_t=1)=\frac{1}{N}\sum_{i=1}^N p_{i,t}. \tag{1}\label{eq:BF1}
\end{equation}
(Here and elsewhere, whenever I reproduce an equation from BF, possibly in my notation, I use their original equation number in parentheses. All other numbered equations follow the numbering of this comment.) Each agent survives between periods with probability $1-\delta$. When an agent dies, a newborn enters the same dynasty and draws an initial belief $z_{i,t}$ uniformly from $[0,1]$. Beliefs follow
\begin{equation}
    p_{i,t+1}=
    \begin{cases*}
        (1-\lambda)p_{i,t}+\lambda s_t & if the agent survives,\\
        z_{i,t} & if a newborn enters dynasty $i$,
    \end{cases*} \tag{5}\label{eq:BF5}
\end{equation}
where $\lambda\in (0,1)$. Combining \eqref{eq:BF1} and \eqref{eq:BF5} and letting $N\to\infty$ gives
\begin{equation}
    P_{t+1}=(1-\delta)\bigl[(1-\lambda)P_t+\lambda s_t\bigr]+\frac{\delta}{2}. \tag{6}\label{eq:BF6}
\end{equation}
Thus, the common signal $s_t$ remains an aggregate shock in the large-population limit.

At a weekly frequency, BF calibrate the model using an annual discount factor of $0.97$, an expected life of $50$ years, and a constant gain $\lambda=\sqrt{\delta/\alpha}$ with $\alpha=1$. They simulate $N=1{,}000{,}000$ agents for $300$ years.

The analysis below draws on the following parts of BF. The self-referential restriction \eqref{eq:BF1} is their Equation (1), in Section III.A. Section III.C specifies the belief process actually simulated---mortality combined with constant-gain learning---and delivers \eqref{eq:BF5} and \eqref{eq:BF6}. Section V presents the market economy, the agent's problem, and the finite-population security prices, ending with Equation (25). Section VI.B derives the pricing relation and wealth recursion under heterogeneous beliefs in Equations (30)--(33). Section VII describes the simulation, while Section VIII and Appendix F examine the wealth distribution and its tail.\footnote{The remaining sections address issues not used in the present analysis. Section III.B studies infinitely lived least-squares learners whose disagreement vanishes. It motivates the change of assumptions made in III.C rather than describing the economy that is simulated; it is also where the mean ergodic theorem noted in Appendix \ref{sec:inventory} appears. Sections III.D and IV characterize the invariant measure and convergence rate of the belief process $P_t$ by itself, without reference to wealth. Section IV.A defines the parameter $\alpha\coloneqq\delta/\lambda^2$ and derives the stationary belief density plotted in the upper-left panel of Figure 5. Section VI.C prices debt and equity, Section VI.D discusses what BF take to drive their results, and Section VIII.B applies the Kelly criterion. None of these enters the wealth statistics at issue here.}

\subsection{Wealth dynamics}\label{subsec:model_dynamics}

BF's wealth variable includes both human and financial wealth. If $a_{i,t}$ denotes the value of Arrow securities brought into date $t$, BF define
\begin{equation}
    W_{i,t}\coloneqq H+a_{i,t}. \tag{14}\label{eq:BF14}
\end{equation}
Thus $W_{i,t}$ measures total wealth available before date-$t$ consumption and purchases of claims on date-$t+1$ states. BF write human wealth as $H_{i,t}$ in their Equation (14); their Equation (24) establishes that it is the same constant for every agent, which I impose from the start. BF set human wealth to
\begin{equation}
    H\coloneqq\frac{\epsilon}{1-\beta(1-\delta)}, \tag{24}\label{eq:BF24}
\end{equation}
where $\epsilon$ denotes the per-period endowment and $\beta$ denotes the discount factor. Log utility gives the consumption policy
\begin{equation}
    c_{i,t}=(1-\beta(1-\delta))W_{i,t}. \tag{19}\label{eq:BF19}
\end{equation}
The agent therefore spends $\beta(1-\delta)W_{i,t}$ on state-contingent claims for the next date. %This saving decision does not produce an additional factor $\beta$ in Equation \eqref{eq:BF33} below, however, because the equilibrium prices of those claims also contain $\beta$.

Arrow securities are in zero net supply, so market clearing gives the accounting identity
\begin{equation}
    \sum_{i=1}^N W_{i,t}=NH \label{eq:aggregate}
\end{equation}
at every date and every population size. Equivalently, cross-sectional mean wealth equals $H$ at every date, whatever the distribution of beliefs and wealth. Identity \eqref{eq:aggregate} is exact rather than asymptotic. Section \ref{subsec:tail_mean} returns to its implications.

In the large-population economy, BF define the high-state probability implied by market prices as
\begin{equation}
    q_{t+1}\coloneqq P_{\imp,t+1}
    =\frac{1}{NH}\sum_{i=1}^N p_{i,t+1}W_{i,t}. \tag{32}\label{eq:BF32}
\end{equation}
Let $j_{t+1}\coloneqq(s_{t+1},x_{t+1})$ denote the joint public-signal and mortality state. Denote the common probability of the mortality outcome by $p(x_{t+1})$. Before imposing market clearing, BF's Equation (21) gives
\begin{equation}
    W_{i,t+1}=
    \begin{cases*}
        \beta\frac{\Pi_{i,t}(j_{t+1})}{Q_t(j_{t+1})}W_{i,t}
        & if the agent survives, \\
        H & if a newborn enters dynasty $i$,
    \end{cases*} \tag{21}\label{eq:BF21}
\end{equation}
where $\Pi_{i,t}(j_{t+1})$ denotes the agent's subjective probability of the joint state and $Q_t(j_{t+1})$ denotes its Arrow-security price. BF assume common and independent beliefs about mortality, so conditional on survival, the relevant terms satisfy
\begin{equation*}
    \Pi_{i,t}(j_{t+1})
    =p(x_{t+1})p_{i,t+1}^{s_{t+1}}
    (1-p_{i,t+1})^{1-s_{t+1}}
\end{equation*}
and their large-population pricing formula becomes
\begin{equation}
    Q_t(j_{t+1})
    =\beta p(x_{t+1})q_{t+1}^{s_{t+1}}
    (1-q_{t+1})^{1-s_{t+1}}. \tag{30}\label{eq:BF30}
\end{equation}
Substitution into Equation \eqref{eq:BF21} cancels both $\beta$ and $p(x_{t+1})$. Define the resulting realized growth factor of total wealth by
\begin{equation}
    G_{i,t+1}
    \coloneqq
    \left(\frac{p_{i,t+1}}{q_{t+1}}\right)^{s_{t+1}}
    \left(\frac{1-p_{i,t+1}}{1-q_{t+1}}\right)^{1-s_{t+1}}. \label{eq:growth}
\end{equation}
BF's Equation (33) then becomes
\begin{equation}
    W_{i,t+1}=
    \begin{cases*}
        G_{i,t+1}W_{i,t} & if the agent survives,\\
        H & if a newborn enters dynasty $i$,
    \end{cases*} \tag{33}\label{eq:BF33}
\end{equation}
Every surviving agent therefore experiences total-wealth growth $p_{i,t+1}/q_{t+1}$ when the signal equals one and $(1-p_{i,t+1})/(1-q_{t+1})$ when it equals zero. Both factors equal one when $p_{i,t+1}=q_{t+1}$, leaving the agent's wealth unchanged. If $p_{i,t+1}>q_{t+1}$, wealth rises when the event occurs and falls otherwise. The difference between these state-contingent growth rates increases with the distance between $p_{i,t+1}$ and $q_{t+1}$. By the accounting identity \eqref{eq:aggregate}, these factors redistribute wealth rather than create it.

BF move from \eqref{eq:BF33} to a claim about the stationary upper tail. They invoke the literature on multiplicative processes with reset and posit
\begin{equation}
    \Pr(W>w)\sim c w^{-\mu} \quad\text{as $w\to\infty$} \tag{39}\label{eq:Pareto}
\end{equation}
for some Pareto exponent $\mu>1$. Appendix F proposes an approximation to $\mu$.

\section{Cross-sectional wealth and dynasty averages}\label{sec:dist}

\subsection{Three different wealth distributions}\label{subsec:dist_estimands}

BF describe the object plotted in Figure 6 as ``the time average of 250 equally spaced samples of the wealth distribution.'' Footnote 22 further states that ``for large $T$, our sample histogram will converge to the ergodic wealth distribution.'' This description admits two interpretations. One is to stack the 250 sampled cross sections into a single pool of $NK$ wealth observations. The resulting object is the mixture of the sampled distributions and is the natural interpretation of a time average \emph{of distributions}. The other is to average each dynasty's wealth across the 250 dates before forming a distribution over the resulting $N$ numbers. The BF replication program implements the second interpretation. It stores log wealth at $K=250$ dates in the $N\times K$ matrix \texttt{LW}. Line 115 then averages along the date dimension by executing \texttt{BW=mean(exp(LW),2)}, returning
\begin{equation}
    \bar W_i\coloneqq\frac{1}{K}\sum_{k=1}^K W_{i,t_k}. \label{eq:timeaverage}
\end{equation}
Every reported statistic---the histogram, the quantiles, the empirical CDF, the Lorenz curve, the Gini coefficient, and the Pareto regression---is computed from these $N$ numbers. The distinction matters because index $i$ does not identify an agent. The occupant of dynasty $i$ dies and is replaced; $\bar W_i$ therefore averages the wealth of the successive agents occupying that dynasty at the sampled dates. What the program reports is the distribution function
\begin{equation}
    \widetilde F_K(w)
    \coloneqq\frac{1}{N}\sum_{i=1}^N
    \boldsymbol{1}\left\{\bar W_i\leq w\right\} \label{eq:timeaveragecdf}
\end{equation}
of dynasty-level average wealth over the simulation window. The empirical wealth distributions used for comparison instead record household wealth at one date.

The program computes neither of the two distributions relevant to BF's question. If the target is wealth at a point in time, a single post-burn-in cross section estimates the history-conditional CDF
\begin{equation}
    \widehat F_{t}(w)
    \coloneqq\frac{1}{N}\sum_{i=1}^N
    \boldsymbol{1}\left\{W_{i,t}\leq w\right\}. \label{eq:crosscdf}
\end{equation}
Large $N$ makes idiosyncratic sampling error small, but it does not average away the common signal history in \eqref{eq:BF6}. If the target is instead the unconditional stationary law, one can pool cross sections over sufficiently separated dates or over independent histories:
\begin{equation}
    \widehat F_K(w)
    \coloneqq\frac{1}{NK}\sum_{k=1}^K\sum_{i=1}^N
    \boldsymbol{1}\left\{W_{i,t_k}\leq w\right\}. \label{eq:pooledcdf}
\end{equation}
Neither \eqref{eq:crosscdf} nor \eqref{eq:pooledcdf} equals \eqref{eq:timeaveragecdf}.

This distinction is substantive and is recognized explicitly in related work. \citet[p.~9, n.~16]{Azarmsa2026} studies a related wealth model with aggregate state dependence. He states that his tail index characterizes the unconditional stationary distribution of a randomly drawn family rather than the cross section at a given date, which depends on the realized aggregate history and may have a different tail index. A Gini coefficient and a Pareto exponent are properties of a specific distribution. BF do not identify the target distribution explicitly. Their program computes the Gini coefficient of $0.7$ and the Pareto exponent of $1.4$ from \eqref{eq:timeaveragecdf}, which equals neither \eqref{eq:crosscdf} nor \eqref{eq:pooledcdf}.

The asymptotics also differ. If BF hold $K=250$ fixed while $T\to\infty$, wider spacing can reduce serial dependence among the observations, but their average still does not become a single wealth draw. If $K\to\infty$ over a fixed 300-year window, \eqref{eq:timeaverage} approaches each dynasty's finite-window time average, which remains dispersed across dynasties. If both $K$ and the horizon diverge and the joint process is ergodic with finite mean, the ergodic theorem gives
\begin{equation}
    \bar W_i\to \E[W]=H, \label{eq:ergodicmean}
\end{equation}
where the second equality is the market-clearing identity \eqref{eq:aggregate}. Then \eqref{eq:timeaveragecdf} collapses to a point mass at $H$ rather than to the stationary cross-sectional distribution, and its Gini coefficient falls to zero. Under ergodicity this limit also does not depend on the realized history. BF's \emph{quasi} nonergodicity implies slow convergence, not a random infinite-horizon limit.

\subsection{Time averaging generates BF's reported numbers}\label{subsec:dist_averaging}

To isolate the estimand, I preserve BF's implemented dynamics and vary only the distributional calculation. Table \ref{tab:estimands} compares three objects calculated from the same simulated history: a terminal cross section, a pool of BF's 250 recorded cross sections, and BF's dynasty-specific average. It reports the first and third objects at three population sizes, including $N=1{,}000{,}000$. Because line 8 of BF's program sets \texttt{rng(2)}, the reported levels correspond to the history underlying Figures 5--7. Averages and standard deviations below are taken across $\numHorizonSeeds$ histories, including the published seed.

\begin{table}[!htb]
    \centering
    \caption{Alternative wealth-distribution estimands.}
    \label{tab:estimands}
    \footnotesize
    \setlength{\tabcolsep}{4pt}
    \begin{tabular}{lrrrrr}
        \toprule
        & Mean$/H$ & Median$/H$ & 99th pct.$/H$ & Gini & Tail exponent \\
        \midrule
        \textit{Terminal cross section} \\
        \quad $N=10{,}000$ & \numXsMean & \numXsMedian & \numXsPninenine & \numXsGini & \numXsTail \\
        \quad $N=100{,}000$ & \numXsLadderMeanHundredK & \numXsLadderMedianHundredK & \numXsLadderPninenineHundredK & \numXsLadderGiniHundredK & \numXsLadderTailHundredK \\
        \quad $N=1{,}000{,}000$ & \numXsLadderMeanMillion & \numXsLadderMedianMillion & \numXsLadderPninenineMillion & \numXsLadderGiniMillion & \numXsLadderTailMillion \\
        \addlinespace
        \textit{Pooled cross sections} \\
        \quad $N=10{,}000$ & \numPooledMean & \numPooledMedian & \numPooledPninenine & \numPooledGini & \numPooledTail \\
        \addlinespace
        \textit{BF dynasty average} \\
        \quad $N=10{,}000$ & \numLadderMeanTenK & \numLadderMedianTenK & \numLadderPninenineTenK & \numLadderGiniTenK & \numLadderTailTenK \\
        \quad $N=100{,}000$ & \numLadderMeanHundredK & \numLadderMedianHundredK & \numLadderPninenineHundredK & \numLadderGiniHundredK & \numLadderTailHundredK \\
        \quad $N=1{,}000{,}000$ & \numLadderMeanMillion & \numLadderMedianMillion & \numLadderPninenineMillion & \numLadderGiniMillion & \numLadderTailMillion \\
        \addlinespace
        \textit{BF, published} & --- & \textit{0.39} & \textit{8.92} & \textit{0.70} & \textit{1.4} \\
        \bottomrule
    \end{tabular}

    \caption*{\footnotesize \textit{Notes:} BF's implemented recursion, parameters, and seed, a 60-year burn-in, and a 300-year weekly horizon. ``Terminal cross section'' is the wealth vector at the final simulated date; ``BF dynasty average'' first averages each dynasty's wealth across the 250 recorded dates, as BF's program does. Both objects are reported at three population sizes, including BF's $N=1{,}000{,}000$. ``Pooled cross sections'' stacks the 250 recorded cross sections before calculating the statistics. It requires the full $N\times K$ matrix and is therefore reported at $N=10{,}000$ only. The tail exponent uses BF's fixed $7\leq\log W\leq12$ regression window (replication lines 16 and 124--135). Market clearing makes cross-sectional mean wealth equal to $H$ exactly; Mean$/H$ should therefore be $1$. The small excess comes from the pricing error in BF's program documented in Appendix \ref{sec:program}. The last row reports the values BF publish on p.~977 and in Figures 6--7. Changing $N$ changes the simulated history even under the same seed because the public signal depends on the simulated beliefs. Rows at different $N$ therefore differ in history as well as in population size.}
\end{table}

The last dynasty-average row reproduces BF's published statistics to the precision at which they report them. It is the relevant comparison because $N=1{,}000{,}000$ is BF's population size and \texttt{rng(2)} is the published seed. The replication returns a Gini coefficient of $\numLadderGiniMillion$ compared with their $0.70$, a median of $\numLadderMedianMillion H$ compared with their $39$ percent, a 99th percentile of $\numLadderPninenineMillion H$ compared with their $892$ percent, and a Pareto exponent of $\numLadderTailMillion$ compared with their $1.4$. The agreement across all four statistics identifies the distribution underlying BF's reported values and confirms that the recursion used throughout this comment is BF's.

None of the cross-sectional rows approximates BF's reported statistics at any population size. The sharpest comparison uses the published seed and $N=1{,}000{,}000$, for which the cross section and dynasty average summarize the same simulation. The cross section has a Gini coefficient of $\numXsLadderGiniMillion$, compared with $\numLadderGiniMillion$ for the dynasty average. Its median is $\numXsLadderMedianMillion H$, compared with $\numLadderMedianMillion H$ for the dynasty average and the $39$ percent BF report.

Time averaging preserves mean wealth, as \eqref{eq:aggregate} requires, but materially changes every other reported statistic. The same identity implies that the Mean$/H$ column should equal $1$ exactly. The observed excess cannot reflect sampling error: BF's program departs from their pricing formula in a way that violates \eqref{eq:aggregate}. Appendix \ref{sec:program} documents the departure. Table \ref{tab:corrected} shows that correcting it restores Mean$/H=1$ while leaving every other statistic in this section essentially unchanged. Pooling the same 250 cross sections, which preserves the unit of observation, changes little. The pooled Gini coefficient is $\numPooledGini$, compared with $\numXsGini$ for a single cross section at the same $N=10{,}000$.

The population-size comparison also supports the use of $N=10{,}000$ in the remaining exercises. Repeating simulations with one million agents across seeds and horizons is computationally expensive. Neither statistic varies systematically with population size: the dynasty-average rows span $\numLadderGiniTenK$ to $\numLadderGiniHundredK$, while the cross-sectional rows span $\numXsLadderGiniHundredK$ to $\numXsLadderGiniTenK$. Each range is of the same order as the standard deviation of $\numDynGiniSD$ that Section \ref{subsec:dist_imply} reports across $\numSweepSeeds$ histories at a fixed $N$. The comparisons below are driven by the common signal history, which a larger $N$ does not average away.

Time averaging also changes the estimated tail exponent, one of BF's four reported distributional statistics. At the common $N=10{,}000$, BF's fixed regression window gives $\numLadderTailTenK$ for the dynasty average, reproducing the reported value of $1.4$. The terminal cross section instead gives $\numXsTail$. Averaging therefore changes the shape of the upper tail rather than merely rescaling wealth. All four statistics BF publish describe the averaged object rather than a cross section.

Neither estimate provides reliable evidence about the asymptotic tail exponent. BF's fixed-window regression gives $\numXsTail$, whereas the Hill estimator---the maximum-likelihood tail estimate computed from the largest observations alone \citep{Hill1975}---gives $\numXsHill$ for the richest one percent of the same cross section. The estimates straddle $\mu=1$. Section \ref{subsec:tail_mean} shows that market clearing pins cross-sectional mean wealth at $H$ exactly. A nondegenerate limiting law therefore cannot have $\mu\leq1$, indicating that these estimates do not measure an asymptotic exponent. A line fitted over one preselected range establishes neither a Pareto law nor a value of $\mu$.

\subsection{BF's statistic is a function of the simulation horizon}\label{subsec:dist_horizon}

\begin{figure}[!htb]
    \centering
    \includegraphics[width=\textwidth]{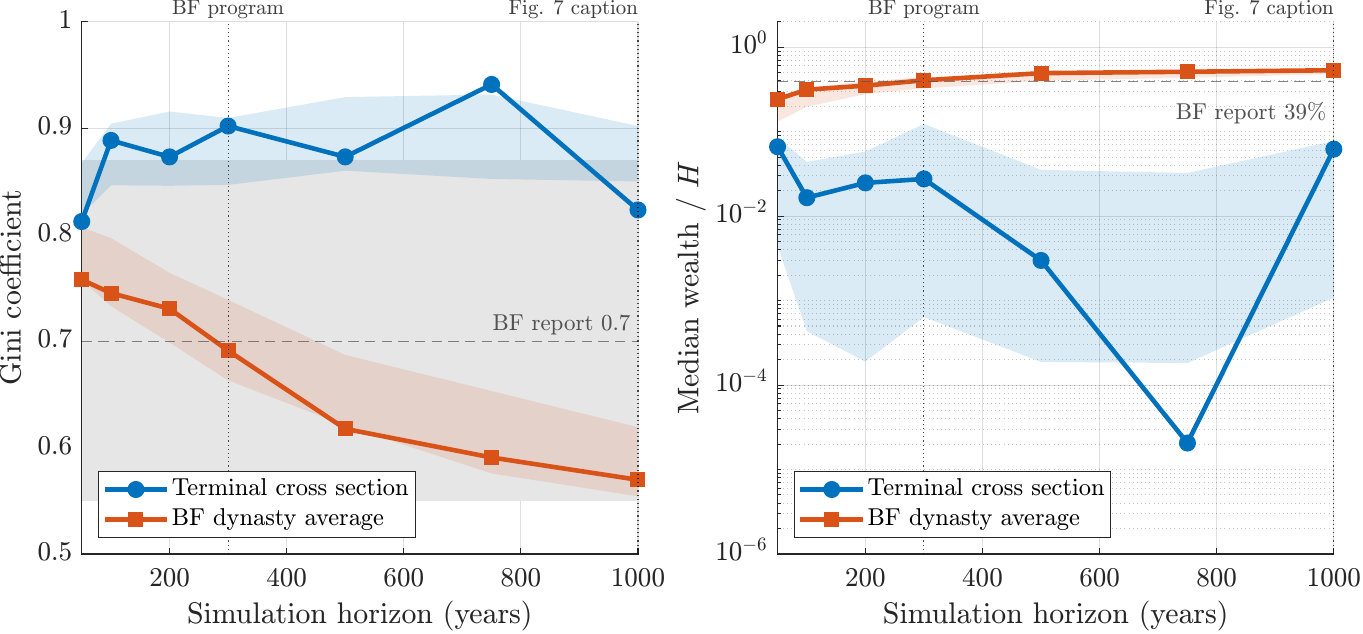}
    \caption{Inequality against the simulation horizon.}
	\caption*{\footnotesize \textit{Notes:} Solid lines use the published history, \texttt{rng(2)}; shaded bands report the mean across $\numHorizonSeeds$ seeds plus and minus one standard deviation. The bands in the right panel are formed in logs to remain positive on the logarithmic axis. The gray band marks the range of national wealth Gini coefficients in BF's Table 1, $0.55$ to $0.87$. The terminal cross section is horizon invariant. BF's dynasty average declines monotonically, crossing their reported $0.7$ just before the 300-year horizon and approaching the bottom of the empirical range at the $\numHorizonLongYears$-year horizon announced in their Figure 7 caption.}
    \label{fig:horizon}
\end{figure}

If \eqref{eq:timeaveragecdf} were a legitimate estimator of a cross-sectional law, a longer sample would increase its precision. Because it is a distribution of time averages, however, a longer sample instead drives it toward the degenerate limit \eqref{eq:ergodicmean}. Figure \ref{fig:horizon} and Table \ref{tab:horizon} report the resulting horizon dependence.

BF sample the wealth distribution at $K=250$ dates that are almost evenly spaced over the post-burn-in window. Line 26 of their program sets \texttt{nums=floor(\allowbreak linspace(1,\allowbreak T,\allowbreak n))} with $T=300\times52$ and \texttt{n}$=250$ from line 23. Consecutive sampled dates therefore lie $62$ or $63$ weeks apart. Line 115 then averages across those dates within each dynasty.\footnote{Two implementation details do not affect the results below. The first stored column is the wealth vector at the end of the burn-in, making the first two samples one week apart rather than $63$. Because the recording loop ends at $T-1$, the last sample falls at week $15{,}537$ rather than at week $15{,}600$.} The horizon experiment retains this design and rescales it. For a horizon of $Y$ years, I hold $K=250$ and the 60-year burn-in fixed. I use BF's recursion and parameters, spreading the $250$ sampled dates evenly over the $52Y$ post-burn-in periods. The spacing between samples grows in proportion to $Y$, while their number remains fixed. The length of the window they span is therefore the only element that changes.

Holding $K$ fixed isolates the effect of the simulation horizon because changing $K$ has little effect on the statistic once $K$ is not small. Over the 300-year window, raising $K$ from $25$ to $250$ changes the dynasty-average Gini coefficient only from $\numKgridTwentyfive$ to $\numDynGini$. Table \ref{tab:horizon} shows that changing the window length changes it by $\numHorizonDynGiniDrop$, more than twenty times as much.

Because BF report a single history, Figure \ref{fig:horizon} presents that history alongside a measure of cross-history dispersion. The solid lines use the published seed, \texttt{rng(2)}. The shaded bands report the mean across $\numHorizonSeeds$ seeds, including the published seed, plus and minus one standard deviation. Each seed uses $N=10{,}000$, as in every other exercise. The horizons are nested within one simulation per seed, holding the public-signal history common across horizons. I include the terminal cross section as a control. Once the burn-in has passed, its law does not depend on the simulation length and should exhibit no horizon trend.

\begin{table}[!htb]
    \centering
    \caption{Inequality against the simulation horizon.}
    \label{tab:horizon}
    \footnotesize
    \setlength{\tabcolsep}{3.5pt}
    \begin{tabular}{lrrrrrrr}
        \toprule
        Horizon (years) & 50 & 100 & 200 & 300 & 500 & 750 & 1{,}000 \\
        \midrule
        \horizonRows
        \bottomrule
    \end{tabular}

    \caption*{\footnotesize \textit{Notes:} BF's implemented recursion and parameters, $K=250$ recorded dates, a 60-year burn-in held fixed at every horizon, and $N=10{,}000$. The level rows report the published history, \texttt{rng(2)}; the standard deviations are taken across $\numHorizonSeeds$ seeds, including the published seed. Horizons are nested within one simulation per seed, holding the public-signal history common across columns. BF's program uses the 300-year column; their Figure 7 caption describes 1,000 years of monthly data.}
\end{table}

Table \ref{tab:horizon} shows three patterns. First, BF's statistic declines monotonically, from $\numHorizonShortDynGini$ at $\numHorizonShortYears$ years to $\numHorizonBaseDynGini$ at $300$ years and $\numHorizonLongDynGini$ at $\numHorizonLongYears$ years. This pattern is consistent with \eqref{eq:ergodicmean}.

Second, the terminal cross-sectional Gini coefficient shows no horizon trend. Under the published history, it varies by $\numHorizonXsGiniRange$ across a twentyfold change in the horizon. This movement reflects variation in a single draw: at any one horizon, the standard deviation across histories reaches $\numHorizonXsGiniSDmax$. The mean across the $\numHorizonSeeds$ histories varies by only $\numHorizonXsGiniMeanRange$ across the seven horizons, without a systematic trend. The contrast identifies the drift as a property of the estimator rather than of the economy.

Third, BF's median rises toward $H$ as the window lengthens. The cross-sectional median, by contrast, remains below $0.07H$ at every horizon in Table \ref{tab:horizon} and exhibits no trend. The decline in the dynasty-average Gini coefficient is not specific to the published seed. The standard deviation of that coefficient across seeds never exceeds $\numHorizonDynGiniSDmax$, while the decline of $\numHorizonDynGiniDrop$ between $\numHorizonShortYears$ and $\numHorizonLongYears$ years is approximately five times the variation across histories. At $\numHorizonLongYears$ years, the mean across seeds is $\numHorizonLongDynGiniMean$ with a standard deviation of $\numHorizonLongDynGiniSD$; the published history yields $\numHorizonLongDynGini$.

The horizon dependence is also inconsistent with BF's documentation. Their Figure 7 caption reports ``1,000 years of simulated monthly data,'' whereas lines 20--22 of the program specify 300 years of weekly data. At the horizon announced in the caption, the same construction would have produced a Gini coefficient of approximately $\numHorizonLongDynGini$. This value lies at the bottom of the range of national wealth Gini coefficients that BF use for comparison and below every country in their table except China in 2008. The construction would also have produced a Pareto exponent of $\numHorizonLongDynTail$ under the published history and $\numHorizonLongDynTailMean$ on average across histories, rather than the reported value of $1.4$. The exponent varies less with the horizon than the Gini coefficient---by approximately two cross-history standard deviations rather than five---so the Gini coefficient provides clearer evidence of horizon dependence. Both numbers published in Figure 7 correspond to the 300-year run performed by the program rather than the $\numHorizonLongYears$-year run described in the caption. This degree of sensitivity to an undocumented and internally inconsistent simulation horizon precludes direct comparison with an empirical cross section.

\subsection{What the cross section implies}\label{subsec:dist_imply}

The corrected measurement confirms BF's qualitative mechanism and implies more inequality, moving the model from the middle of the empirical range to its upper edge. This direction is consistent with complete markets in the present value of endowment claims, which allow agents to take large leveraged positions. It does not make the choice of distributional estimand immaterial: BF compare their reported statistics with empirical cross sections. With BF's parameters, seed, and both program corrections from Appendix \ref{sec:program}, the terminal cross section has a Gini coefficient of $\numBothGini$. Under BF's implemented recursion, which Table \ref{tab:estimands} and the rest of this comment use, the same cross section gives $\numXsGini$. The two program errors therefore account for little of the gap with BF's $0.7$. Across $\numSweepSeeds$ histories, the corrected cross-sectional Gini coefficient averages $\numCorrGiniMean$, has a standard deviation of $\numCorrGiniSD$, and ranges from $\numCorrGiniMin$ to $\numCorrGiniMax$. BF's Table 1 reports national wealth Gini coefficients between $0.55$ and $0.87$; the two largest entries for 2019 are Switzerland at $0.87$ and the United States at $0.85$.\footnote{On p.~976, BF describe the same table as varying ``between a low of $0.55$ for China in 2008 and a high of $0.85$ for the United States in 2019,'' which omits the $0.87$ entry for Switzerland in their own Table 1. I use the table.} The model's cross-sectional inequality lies at the upper boundary of that range and exceeds its maximum on average, rather than lying near its middle as BF's $0.7$ does. Assessing the model's empirical fit requires a separate criterion. The relevant result here is that the cross-sectional value differs substantially from the value BF report.

The discrepancy is larger for median wealth. \citet[p.~977]{BouchaudFarmer2023} state ``a person at the 50th percentile of the wealth distribution is a net borrower who has total wealth equal to 39\% of human wealth.'' In the cross section, median total wealth equals $\numBothMedian H$ (Table \ref{tab:corrected}, last row), or approximately two percent of human wealth. The model therefore implies that half the population is nearly fully leveraged against future labor income at any date. This strong and testable prediction differs from the one BF report.

Cross-history variation is also material, but BF do not report it. Their footnote 22 states that unreported experiments with other draws suggest limited variability. Neither the article nor the archive, however, presents a Monte Carlo summary. Because the public signal remains an aggregate shock as $N\to\infty$, large $N$ removes idiosyncratic sampling error but not cross-history variation. The dynasty-average Gini coefficient ranges from $\numDynGiniMin$ to $\numDynGiniMax$ over $\numSweepSeeds$ histories, with a standard deviation of $\numDynGiniSD$, while the fixed-window exponent ranges from $\numDynTailMin$ to $\numDynTailMax$.

\section{Pareto-tail analysis}\label{sec:tail}

Quite apart from the measurement problem discussed in Section \ref{sec:dist}, a Pareto upper tail is a recurring claimed result throughout BF. The Introduction states that the model generates a Pareto-tailed wealth distribution and reproduces the empirical value of its exponent \citep[p.~950]{BouchaudFarmer2023}. The literature review says that the asymptotic wealth distribution displays a Pareto tail \citep[p.~952]{BouchaudFarmer2023}. Section VIII says that Figure 7 ``reveals a power-law tail'' with exponent $1.4$, reports the same exponent under two alternative calibrations, and claims a nondegenerate value greater than one as mortality vanishes \citep[pp.~976, 978--980]{BouchaudFarmer2023}. The Conclusion again states that the coupled wealth-belief dynamics lead to a fat-tailed wealth distribution \citep[p.~982]{BouchaudFarmer2023}. The Pareto tail is therefore part of the paper's quantitative and theoretical contribution, not an incidental feature of an appendix.

Appendix F provides the paper's dedicated analytical support for these claims. Immediately after positing the Pareto tail in Equation (39), BF refer to Appendix F for an approximate calculation of the Pareto exponent $\mu$ and the conclusion that $\mu>1$ whenever $\delta>0$; footnote 28 uses its Equation (F5) for the small-mortality limit. Appendix F is explicitly a simplified analysis, so it need not reproduce the fully articulated equilibrium exactly. The criticism here is instead that its multiplicative-reset kernel does not describe the BF wealth process and that its formulas do not support the conclusions for which BF invoke them. The remainder of this section establishes these points; Appendix \ref{sec:tail_details} provides the supporting calculations.

\subsection{Multiplicative-reset kernel is not the BF transition}\label{subsec:tail_kernel}

Consider the multiplicative process with reset
\begin{equation}
    W'=
    \begin{cases*}
        GW & with probability $1-\delta$,\\
        1 & with probability $\delta$,
    \end{cases*} \tag{F1}\label{eq:iidreset}
\end{equation}
where $G>0$ is independent of current wealth and has the same distribution over time. BF define the one-period net wealth growth rate by $\eta\coloneqq G-1$. Appendix F assumes that $\eta$ has mean zero and variance $\sigma^2$. Equation \eqref{eq:iidreset} concerns the marginal evolution of one wealth process and does not, by itself, require different dynasties to draw independent multipliers $G$. For example, suppose that every dynasty faces the same \iid public multiplier but experiences an independent reset. A randomly selected dynasty still follows \eqref{eq:iidreset}. The common multiplier nevertheless induces dependence across dynasties. A realized cross section therefore need not reproduce the dynasty's unconditional stationary law.

If the stationary upper tail of \eqref{eq:iidreset} satisfies \eqref{eq:Pareto}, the tail-balance heuristic gives
\begin{equation}
    (1-\delta)\E[G^\mu]=1, \tag{F3}\label{eq:iidexponent}
\end{equation}
which is BF's Equation (F3); the same condition appears in \citet{ManrubiaZanette1999}. \citet[p.~1815]{BeareToda2022ECMA} derive this condition together with the spectral-radius condition that replaces \eqref{eq:iidexponent} when the distribution of $G$ is driven by an exogenous finite-state Markov chain. These results characterize the stationary law of one multiplicative process. Identifying that unconditional law from one large cross section additionally requires a suitable cross-sectional law of large numbers and explicit treatment of any common history.

This distinction applies directly to the BF economy. Dynasties face idiosyncratic deaths and newborn beliefs, but every survivor responds to the same public signal. A date-$t$ cross section therefore estimates a distribution conditional on one realized public history. Taking $N\to\infty$ averages mortality and newborn beliefs. It does not average the common sequence $s^t=(s_1,\ldots,s_t)$ or turn the cross section into independent draws from an unconditional stationary law.

More fundamentally, the analysis of BF does not supply the one-dynasty transition kernel assumed in their Equation (F2). In their model,
\begin{equation*}
    G_{i,t+1}=G\left(p_{i,t+1},q_{t+1},s_{t+1}\right),
\end{equation*}
where $G(p,q,s)\coloneqq (p/q)^s[(1-p)/(1-q)]^{1-s}$ is the wealth-growth factor in \eqref{eq:growth}. BF emphasize after their Equation (33) that belief and wealth become strongly coupled. Because $q_{t+1}$ depends on the joint cross-sectional distribution of beliefs and wealth, the conditional distribution of $G_{i,t+1}$ generally depends on current wealth. The product measure $\varrho(w)p(\eta)$ in BF's Equation (F2) therefore does not represent the BF transition.

An application of the Markov extension would require a state containing at least the dynasty's belief and the aggregate wealth-belief distribution that determines $q_{t+1}$. BF neither construct that state nor derive its tail equation. Appendix F instead analyzes a different stochastic process and uses it as a heuristic for the equilibrium. Although common shocks need not rule out a Pareto tail, neither the moment condition \eqref{eq:iidexponent} nor its cross-sectional interpretation follows from BF's public-signal economy.

\subsection{Appendix F formulas at BF's calibration}\label{subsec:tail_calibrated}

Appendix F uses \eqref{eq:iidexponent} to obtain
\begin{equation}
    \mu\approx \frac{1}{2}\left(1+\sqrt{1+\frac{8\delta}{\sigma^2}}\right). \tag{F4}\label{eq:BFF4}
\end{equation}
Noting that returns are persistent, BF assume in a simplified calculation that $\eta$ remains constant for $\lambda^{-1}$ periods and replace \eqref{eq:BFF4} by
\begin{equation}
    \mu_{\mathrm{BF}}
    \approx \frac{1}{2}\left(1+\sqrt{1+\frac{8\delta\lambda}{\bar\sigma^2}}\right),
    \qquad \bar\sigma^2\coloneqq \E[\sigma^2(t)]. \tag{F5}\label{eq:BFF5}
\end{equation}
BF derive \eqref{eq:BFF4}, in their words, ``in the limit when $\delta$ and $\sigma^2$ are small,'' and obtain \eqref{eq:BFF5} from a simplified version of the same expansion. Both can be evaluated directly at the calibrated parameters. The quantity $\bar\sigma^2$ is the Equation (38) moment of the model's net return $\eta$, defined by $\eta\coloneqq G_{i,t+1}-1$ using \eqref{eq:growth}; the calibration fixes $\delta$ and $\lambda$. Table \ref{tab:appendixF} reports the result of measuring that moment in the simulated economy and substituting it into the formulas.

\begin{table}[!htb]
    \centering
    \caption{Appendix F evaluated at BF's parameters.}
    \label{tab:appendixF}
    \small
    \begin{tabular}{lr}
        \toprule
        Quantity & Value \\
        \midrule
        $\delta$ (weekly mortality) & $3.85\times10^{-4}$ \\
        $\lambda$ (belief gain) & $1.96\times10^{-2}$ \\
        $\bar\sigma^2=\E[\eta^2]$, BF Equation (38) & $\numSigmaBarSq$ \\
        \addlinespace
        $\mu$ from (F4) & $\numMuFfour$ \\
        $\mu$ from (F5) & $\numMuFfive$ \\
        \addlinespace
        \textit{BF's reported exponent} & \textit{1.4} \\
        \bottomrule
    \end{tabular}

    \caption*{\footnotesize \textit{Notes:} Corrected dynamics, BF's parameters, $N=10{,}000$, a 300-year weekly horizon after a 60-year burn-in. The moment $\bar\sigma^2$ is the time average of the cross-sectional mean of the conditional second moment of $\eta$ over surviving agents, as BF define it in Equation (38). BF's implemented recursion gives $\bar\sigma^2=\numSigmaBarSqBF$, $\mu=\numMuFfourBF$ from (F4), and $\mu=\numMuFfiveBF$ from (F5).}
\end{table}

Appendix F does not support BF's reported exponent at the model's calibration. An approximation need not reproduce a simulated estimate exactly. Here, however, the formulas return $\mu=\numMuFfour$ from (F4) and $\mu=\numMuFfive$ from (F5) when evaluated at BF's parameters using the model's second moment. BF report $1.4$ from a regression on simulated data and describe Appendix F as showing ``how $\mu$ can be approximately computed.'' The approximations differ from the reported value by approximately $0.4$ and place $\mu$ near the boundary $\mu=1$, at which the mean ceases to exist.

The small radicands explain the result: $8\delta/\bar\sigma^2=\numArgumentFfour$ in (F4) and $8\delta\lambda/\bar\sigma^2=\numArgumentFfive$ in (F5). Across $\numSweepSeeds$ simulated histories, (F4) returns $\mu$ between $\numMuFfourSweepMin$ and $\numMuFfourSweepMax$, with still smaller values from (F5). Appendix \ref{sec:tail_details} documents the underlying moments and their cross-history variation, evaluates the small-mortality argument, and shows that (F5) does not follow from its motivating block-constant approximation.

These calculations do not imply that the model's cross section has $\mu\approx1$. As Section \ref{subsec:tail_kernel} shows, Appendix F does not characterize the model's transition, so its formulas cannot be interpreted as estimates of the cross-sectional exponent. The calculations instead show that BF's analytical arguments for $\mu\approx1.4$, for $\mu>1$ whenever $\delta>0$, and for a nondegenerate small-mortality limit are not supported when the formulas are evaluated at the model's parameters.

\subsection{Market clearing already implies a finite mean}\label{subsec:tail_mean}

BF present $\mu>1$ as a substantive implication of Appendix F. Footnote 27 uses this result to place their economy in the empirically relevant class $1<\mu<2$. The identity \eqref{eq:aggregate} establishes the conclusion directly and independently of Appendix F. Arrow securities are in zero net supply; cross-sectional mean wealth therefore equals $H$ exactly at every date and for every population size. If the sequence of cross-sectional distributions has a nondegenerate limit as $N\to\infty$ and is uniformly integrable, that limit has mean $H$ and cannot be a Pareto law with $\mu\leq1$.

The accounting identity both supports BF's finite-mean conclusion and shows that Appendix F is unnecessary for that conclusion. The finite mean follows from an equilibrium accounting identity rather than a tail-balance calculation. Appendix F would instead need to establish the value of $\mu$ and the asymptotic form of the tail, which it does not.

\section{Conclusion}\label{sec:conclusion}

The corrected calculation confirms BF's central qualitative result: self-referential beliefs coupled with trading generate substantial, indeed extreme, wealth inequality. Their reported Gini coefficient and related statistics, however, describe dynasty-level time averages, not a wealth cross section. Across histories, the corrected cross-sectional Gini coefficient averages $\numCorrGiniMean$, placing the model at the upper end rather than the middle of BF's empirical comparison range. Median total wealth is approximately two percent of human wealth. The closer empirical fit reported by BF is attributable to the time-averaging procedure.

These results do not rule out a Pareto tail generated by self-referential beliefs; the corrected cross sections display substantial upper-tail concentration. Appendix F, however, characterizes a different stochastic process and establishes neither an asymptotic Pareto law nor its exponent. The self-referential belief process may therefore generate substantial inequality and tail concentration, but the reported quantitative fit and claimed exponent do not follow from BF's analysis.

\printbibliography

\newpage
\appendix

\begin{center}
	{\Large \textbf{Online Appendix (Not for publication)}}
\end{center}

\section{Implementation of replication program}\label{sec:program}

BF provide MATLAB programs for every figure in the replication archive. Two operations in that program depart from the economy stated in the article. I document them here rather than in the main text because, as Table \ref{tab:corrected} shows, correcting both changes the terminal cross section very little. The programming errors are worth recording, but they are not the source of BF's reported numbers.

\subsection{Mortality normalization and aggregate wealth}\label{subsec:program_normalization}

Let $x_i\in\set{0,1}$ indicate survival in a simulated period, and define
\begin{equation*}
    N_x\coloneqq\sum_{i=1}^N x_i,
    \qquad
    A_1\coloneqq\sum_{i=1}^N x_i p_iW_i,
    \qquad
    A_0\coloneqq\sum_{i=1}^N x_i(1-p_i)W_i.
\end{equation*}
BF's own finite-population pricing formula is their Equation (25), which under their independence assumption for mortality becomes their Equation (30):
\begin{equation}
    Q(j'\mid j)=\beta p(x')\frac{\sum_{i=1}^N p_i(s')W_i(j)x_i'}{N(j')H},
    \qquad N(j')\coloneqq\sum_i x_i', \tag{30}\label{eq:BF30fin}
\end{equation}
where $N(j')$ is the realized number of survivors. After canceling the common discount and mortality-state factors in wealth returns, \eqref{eq:BF30fin} gives the signal-price factors
\begin{equation}
    q_1=\frac{A_1}{N_xH},
    \qquad
    q_0=\frac{A_0}{N_xH}. \label{eq:finiteprices}
\end{equation}
BF's large-population Equation (32) instead integrates out mortality and uses $q=\sum_i p_iW_i/(NH)$.

The program uses neither formula. Lines 63 and 90 calculate \texttt{nx=sum(x)}, which is $N_x$, and never use it. Lines 67--68 and 96--97 calculate the high-signal factor
\begin{equation}
    q_{1,\code}\coloneqq\frac{A_1}{NH}, \label{eq:qcode}
\end{equation}
and then set the low-signal factor to $1-q_{1,\code}$ rather than to $A_0/(N_xH)$. The program therefore keeps realized survival indicators in the numerator of \eqref{eq:BF30fin} while retaining the large-population denominator of (32), and forces two factors conditional on different mortality states to sum to one. As $N\to\infty$, \eqref{eq:qcode} converges to $(1-\delta)q$, not $q$.

This mismatch violates \eqref{eq:aggregate} directly. When the signal equals one, lines 99--101 give surviving agents the return $p_i/q_{1,\code}$, and \eqref{eq:qcode} then implies
\begin{equation*}
    \sum_{i=1}^N x_i\frac{p_i}{q_{1,\code}}W_i
    =\frac{A_1}{A_1/(NH)}=NH.
\end{equation*}
The program next gives each of the $N-N_x$ newborns wealth $H$, so total wealth after a high signal equals $(2N-N_x)H$, which exceeds $NH$ whenever a death occurs. Under \eqref{eq:finiteprices}, survivors collectively hold exactly $N_xH$ in either signal state, and newborn wealth restores total wealth to $N_xH+(N-N_x)H=NH$. The violation is visible in the Mean$/H$ column of Table \ref{tab:corrected}: BF's normalization gives $\numBFMean$ where market clearing requires exactly $\numBothMean$.

Because the program forces the two factors to sum to one, the low-signal factor can also turn negative. Note that under \eqref{eq:finiteprices} the high-signal factor may itself exceed one without difficulty, since $q_1$ and $q_0$ are state prices and not probabilities of complementary events; the failure comes from the imposed complementarity, not from $q_1>1$ as such. When that happens, the \texttt{real} wrapper on line 99 silently replaces the negative payoff denominator by its absolute value, since $\operatorname{Re}\log x=\log\abs{x}$ for negative real $x$. Across $\numSweepSeeds$ 300-year simulations, the high-signal factor reached or exceeded one an average of $\numHighFactorCount$ times per history, and the low signal activated this absolute-value payoff an average of $\numAbsPayoffCount$ times.

\subsection{Signal timing}\label{subsec:program_timing}

Each iteration also departs from the model's timing. Line 92 draws a signal from the current true probability. Line 94 uses that signal to update beliefs according to \eqref{eq:BF5}. Lines 96--98 calculate the implied probability from the updated beliefs. Finally, lines 99--101 use the same signal to select the high- or low-state return based on those updated beliefs.

Under BF's model, $p_{i,t}$ forecasts $s_t$. After the public observes $s_t$, Equation \eqref{eq:BF5} produces $p_{i,t+1}$, which forecasts $s_{t+1}$. Agents must settle securities paying on $s_t$ with beliefs and prices formed before they observe $s_t$. The program instead updates beliefs with $s_t$ and immediately applies the return on $s_t$ to those post-outcome beliefs, so the program rewards an agent's belief revision with the return on information that the agent has already seen. A correct implementation settles the current contingent claim with the pre-outcome beliefs, then updates beliefs and prices the next claim.

\subsection{Quantitative effect of the corrections}\label{subsec:program_corrected}

Table \ref{tab:corrected} compares four simulations: BF's program, the mortality-price correction alone, the timing correction alone, and both corrections. The price-corrected simulations use \eqref{eq:finiteprices}, which enforces \eqref{eq:aggregate} exactly and converges to BF's large-population recursion. The timing-corrected simulations settle the contingent claim before updating beliefs. All four use common uniform random draws, BF's parameters and seed, a 60-year burn-in, a subsequent 300-year weekly horizon, and $N=10{,}000$, and every statistic comes from one terminal cross section.\footnote{Because the aggregate signal depends on the simulated beliefs, changing $N$ changes the signal history even under the same seed. I therefore use the population-size exercise as a robustness check, not as a numerical convergence proof.}

\begin{table}[!htb]
    \centering
    \caption{Correcting the program: terminal wealth cross section.}
    \label{tab:corrected}
    \small
    \setlength{\tabcolsep}{4pt}
    \begin{tabular}{lrrrrrr}
        \toprule
        & Mean$/H$ & Median$/H$ & 99th pct.$/H$ & Gini
        & BF tail & Hill tail \\
        \midrule
        \correctionRows
        \bottomrule
    \end{tabular}

    \caption*{\footnotesize \textit{Notes:} BF's parameters and seed, $N=10{,}000$, a 60-year burn-in, and one terminal cross section after a subsequent 300-year weekly simulation. The variants use common uniform random draws. Market clearing requires Mean$/H=1$ exactly. ``BF tail'' is minus the fitted slope from BF's log-survival regression over $7\leq\log W\leq12$; ``Hill tail'' uses the richest one percent.}
\end{table}

The mortality correction restores market clearing exactly but barely changes the terminal distribution, because $\delta$ is of order $10^{-4}$ per week; the Gini coefficient is unchanged to three decimals. The timing correction matters more. It lowers median wealth from $\numXsMedian H$ to $\numBothMedian H$, raises the Gini coefficient from $\numXsGini$ to $\numBothGini$, and lowers both tail estimates. Correcting the program therefore leaves the extreme inequality and the heavy-tailed appearance intact, and if anything sharpens them. The program does not implement BF's equilibrium, but these errors are not what makes their reported inequality look realistic.

Two further observations bear on the tail evidence. First, BF hard-code the tail window: line 16 sets \texttt{LB=7; UB=12}, lines 130--135 fit the regression within those bounds, and Figure 7 marks the same endpoints. BF neither explain this choice nor report sensitivity to other windows. Second, alternative tail observations give materially different exponents: across $\numSweepSeeds$ histories, BF's fixed-window estimate on the corrected cross section averages $\numCorrTailMean$ with standard deviation $\numCorrTailSD$, while the top-one-percent Hill estimate averages $\numCorrHillMean$ with standard deviation $\numCorrHillSD$. A fitted line over one preselected range does not establish an asymptotic Pareto law or identify its exponent.

\section{Further analysis of Appendix F}\label{sec:tail_details}

This appendix records the supporting calculations summarized in Section \ref{sec:tail}. It documents the disappearance of idiosyncratic return variation with age, additional features of the moments entering (F4) and (F5), and the block-constant approximation motivating (F5).

\subsection{Idiosyncratic return variation disappears within a few years of birth}\label{subsec:tail_homogenize}

Appendix F notes that \eqref{eq:iidreset} does not apply directly, because ``agent $i$ will consistently make or lose money as long as the sign of $P_i(t)-P(t)$ is constant---i.e., during a time $\sim\lambda^{-1}$''. This observation motivates \eqref{eq:BFF5}, which treats the net return $\eta$ as constant over such an interval and renewed thereafter. BF define $\alpha\coloneqq\delta/\lambda^2$ and calibrate it to $1$, implying $\lambda=\sqrt\delta$. The resulting belief memory time is $\lambda^{-1}=\numInverseLambdaWeeks$ weeks, or $\numInverseLambdaYears$ years, compared with an expected life of $50$ years. Because \eqref{eq:BF5} is an affine recursion driven by the common signal, belief differences between any two surviving agents contract by the factor $1-\lambda$ each period regardless of their initial beliefs. Two agents who have both survived $n$ periods differ by $(1-\lambda)^n$ times their initial difference.

Table \ref{tab:homogenize} measures the consequence in the simulated economy.

\begin{table}[!htb]
    \centering
    \caption{Cross-sectional dispersion of beliefs by age.}
    \label{tab:homogenize}
    \footnotesize
    \begin{tabular}{lcccc}
        \toprule
        Age & Under 1 year & 1 to 5 years & 5 to 20 years & Over 20 years \\
        \midrule
        Standard deviation of beliefs
        & $\numBeliefSDyoung$ & $\numBeliefSDone$ & $\numBeliefSDfive$ & $\numBeliefSDtwenty$ \\
        \bottomrule
    \end{tabular}

    \caption*{\footnotesize \textit{Notes:} Corrected dynamics, BF's parameters, $N=10{,}000$, averaged over dates in a 300-year weekly simulation after a 60-year burn-in. Newborn beliefs are uniform on $[0,1]$, whose standard deviation is $0.289$.}
\end{table}

Agents older than five years agree to about three decimal places, while agents older than twenty years agree to ten decimal places. By \eqref{eq:growth}, agents holding the same belief earn the same return in every state. Their relative wealth therefore remains constant. In BF's calibration, essentially all cross-sectional wealth dispersion is created during the first few years of an agent's fifty-year life. Thereafter, survivors' log wealth differences remain constant and only the common factor changes.

This provides a second, independent reason why \eqref{eq:iidreset} does not characterize the calibrated economy. In addition to the common shock across dynasties, the idiosyncratic component of returns vanishes with age rather than being renewed every $\lambda^{-1}$ periods. A tail argument for this economy would need to characterize the joint distribution of age and birth-cohort luck along a realized public history. Appendix F contains no such argument.

\subsection{Calibrated moments and small mortality}\label{subsec:tail_moments}

Table \ref{tab:appendixF-details} reports the moments and transformations used to evaluate (F4), (F5), and the small-mortality argument.

\begin{table}[!htb]
    \centering
    \caption{Details of Appendix F evaluated at BF's parameters.}
    \label{tab:appendixF-details}
    \small
    \begin{tabular}{lr}
        \toprule
        Quantity & Value \\
        \midrule
        $\delta$ (weekly mortality) & $3.85\times10^{-4}$ \\
        $\lambda$ (belief gain) & $1.96\times10^{-2}$ \\
        $\E[\eta]$, BF Equation (37) & $\numMeanEta$ \\
        $\bar\sigma^2=\E[\eta^2]$, BF Equation (38) & $\numSigmaBarSq$ \\
        $\delta/\bar\sigma^2$ (footnote 28 assumes $\to 1$) & $\numDeltaOverSigmaSq$ \\
        $8\delta/\bar\sigma^2$ (the radicand in (F4)) & $\numArgumentFfour$ \\
        $8\delta\lambda/\bar\sigma^2$ (the radicand in (F5)) & $\numArgumentFfive$ \\
        \addlinespace
        $\mu$ from (F4) & $\numMuFfour$ \\
        $\mu$ from (F5) & $\numMuFfive$ \\
        $\mu$ from the block-constant condition \eqref{eq:blockmu} & $\numMuBlock$ \\
        \addlinespace
        \textit{BF's reported exponent} & \textit{1.4} \\
        \bottomrule
    \end{tabular}

    \caption*{\footnotesize \textit{Notes:} Corrected dynamics, BF's parameters, $N=10{,}000$, a 300-year weekly horizon after a 60-year burn-in. The moments are time averages of the cross-sectional means of the conditional first and second moments of $\eta$ over surviving agents, exactly as BF define them in Equations (37) and (38). BF's implemented recursion gives $\bar\sigma^2=\numSigmaBarSqBF$, $\mu=\numMuFfourBF$ from (F4), and $\mu=\numMuFfiveBF$ from (F5).}
\end{table}

The two radicands explain why (F4) and (F5) return values near one. Weekly mortality implies $\delta=3.85\times10^{-4}$, while $\bar\sigma^2$ is approximately $\numSigmaBarRatio$ times larger. Consequently, $8\delta/\bar\sigma^2=\numArgumentFfour$ and $8\delta\lambda/\bar\sigma^2=\numArgumentFfive$ are both far below one. Because $\frac12(1+\sqrt{1+x})\to1$ as $x\to0$, the formulas provide little quantitative discrimination at this calibration. The discrepancy can be quantified by inverting them. Setting $\frac12(1+\sqrt{1+x})=1.4$ gives $x=2.24$. At BF's mortality rate, (F4) returns $1.4$ only at $\bar\sigma^2=\numSigmaBarSqNeededFfour$, while (F5) does so only at $\bar\sigma^2=\numSigmaBarSqNeededFfive$. Across the $\numSweepSeeds$ simulated histories, the smallest value of $\bar\sigma^2$ is $\numSigmaBarSqMin$.

The moment $\bar\sigma^2$ also varies substantially across histories. It is a time average of a quantity containing terms that diverge whenever the wealth-weighted mean belief $q_t$ approaches zero or one. A small number of dates therefore dominate the average. Across $\numSweepSeeds$ histories, the top one percent of dates contribute up to $\numTopDateShare$ percent of the total, while $\bar\sigma^2$ ranges from $\numSigmaBarSqMin$ to $\numSigmaBarSqMax$. Consequently, there is no history-invariant value at which to evaluate the formulas. Nevertheless, the finding that the formulas yield $\mu$ close to one is robust across histories because the map from $\bar\sigma^2$ to $\mu$ is flat over the observed range. Over the entire range, (F4) returns $\mu$ between $\numMuFfourSweepMin$ and $\numMuFfourSweepMax$, with still smaller values from (F5).

The calibrated moments do not support the limiting argument in footnote 28. BF obtain a nondegenerate small-mortality limit by taking $\delta\to0$ at fixed $\lambda$ in (F5) and asserting that ``the limit of $\delta/\sigma^2$ converges to one.'' This assumption yields $\mu\to\frac12(1+\sqrt{1+8\lambda})=1.23$ at $\lambda=0.14$. The measured ratio at their calibration is $\delta/\bar\sigma^2=\numDeltaOverSigmaSq$ and is smaller on histories with a larger $\bar\sigma^2$. Their conclusion depends entirely on a normalizing constant that differs from its simulated value by a factor of $\numSigmaBarRatio$. Using the value produced by the model, (F5) gives $\mu\to1$, a degenerate limit rather than the stated nondegenerate limit.

\subsection{Persistence correction does not follow from block approximation}\label{subsec:tail_F5}

Equation \eqref{eq:BFF5} also does not follow from the block-constant approximation that motivates it. To formalize BF's statement, let $L$ be an integer of order $\lambda^{-1}$ and suppose that $\eta$ remains constant within a block and is independent across blocks. Conditional on survival throughout a block, wealth grows by $(1+\eta)^L$, while the probability of surviving the block equals $(1-\delta)^L$. The tail condition at block boundaries therefore becomes
\begin{equation*}
    (1-\delta)^L\E\left[(1+\eta)^{\mu L}\right]=1.
\end{equation*}
For small $\delta$ and $\bar\sigma^2$, the same second-order expansion that BF use gives $(1-\delta)^L=1-\delta L+o(\delta L)$ and $\E[(1+\eta)^{\mu L}]=1+\tfrac12\mu L(\mu L-1)\bar\sigma^2+o(\bar\sigma^2)$, so that
\begin{equation}
    \mu L(\mu L-1)\bar\sigma^2\approx 2\delta L \implies
    \mu_{\mathrm{block}}
    \approx \frac{1}{2L}
    \left(1+\sqrt{1+\frac{8\delta L}{\bar\sigma^2}}\right). \label{eq:blockmu}
\end{equation}
Setting $L=1$ recovers \eqref{eq:BFF4}, as required.

Equations \eqref{eq:blockmu} and \eqref{eq:BFF5} agree at leading order in the opposite regime. When $8\delta\lambda/\bar\sigma^2\gg1$, both reduce to $\mu\approx\sqrt{2\delta\lambda/\bar\sigma^2}$. In this regime, replacing $\sigma^2$ by the effective per-period log variance $\bar\sigma^2/\lambda$, as \eqref{eq:BFF5} does, is a defensible coarse-graining when the multiplier dominates. At BF's calibration, however, $8\delta\lambda/\bar\sigma^2=\numArgumentFfive$. The additive constants therefore determine the numerical result: \eqref{eq:BFF5} returns $\numMuFfive$, whereas \eqref{eq:blockmu} returns $\numMuBlock$.

Persistence affects both the exponent on the block multiplier and the probability of surviving the block. BF modify the variance term but leave the remaining terms as though the process drew a new return every period. Consequently, \eqref{eq:BFF5} imposes $\mu\geq1$ algebraically for every parameter configuration. This restriction underlies BF's assertion that the wealth distribution ``always has a finite mean when $\delta>0$'' without further argument. Equation \eqref{eq:blockmu} imposes no such restriction. At BF's calibration, it returns $\numMuBlock$. Under the scaling $\delta/\bar\sigma^2\to1$ assumed in footnote 28, it returns $\tfrac{\lambda}{2}(1+\sqrt{1+8/\lambda})=0.60$ at $\lambda=0.14$, rather than BF's $1.23$.

Equation \eqref{eq:blockmu} is not proposed as a Pareto-exponent formula for the BF economy. The block-constant process supplies only a heuristic approximation. It serves to show that, even under BF's simplification, the finiteness of the mean and the value of the small-mortality limit arise from the particular algebraic form of (F5) rather than from the process it approximates.

\section{Inventory of issues}\label{sec:inventory}

Tables \ref{tab:inventory-central} and \ref{tab:inventory-additional} distinguish central issues from additional ones. The ``Consequence'' column states the implication for BF's analysis. An issue is central when the argument or output named in the ``Location'' column does not establish the conclusion drawn from it. Additional issues materially weaken an inference without alone disposing of a principal result. This classification does not rule out a related result from a different proof or simulation. For concision, I omit typographical and numerically negligible discrepancies.

\begingroup
\footnotesize
\setlength{\tabcolsep}{3.5pt}
\begin{longtable}{@{}c L{0.16\textwidth} L{0.43\textwidth} L{0.29\textwidth}@{}}
    \caption{Inventory of central issues in \citet{BouchaudFarmer2023}.}\label{tab:inventory-central}\\
    \toprule
    No. & Location & Issue or unsupported step & Consequence \\
    \midrule
    \endfirsthead

    \multicolumn{4}{l}{\tablename\ \thetable\ (continued)}\\
    \toprule
    No. & Location & Issue or unsupported step & Consequence \\
    \midrule
    \endhead

    \midrule
    \multicolumn{4}{r}{Continued on next page}\\
    \endfoot

    \bottomrule
    \endlastfoot

    1 & Figures 5--7 and p.~977; replication line 115 & The program uses $\bar W_i\coloneqq K^{-1}\sum_k W_{i,t_k}$ as if it were wealth in a cross section, pooling the different agents who successively occupy one dynasty. Every reported distributional statistic is computed from this object. & BF's Gini coefficient, Lorenz curve, quantiles, and tail regression do not describe cross-sectional wealth. \\

    2 & Page 976, footnote 22 & BF claim that the histogram based on 250 dates converges to the ergodic wealth distribution. Under ergodicity and a finite mean, dynasty time averages converge to the common mean $H$, so the histogram collapses to a point mass. & The proposed convergence argument implies that the reported Gini falls with the horizon rather than converging to the stationary cross-sectional Gini. \\

    3 & Figure 7 caption against replication lines 20--22 & The caption describes 1,000 years of monthly data; the program simulates 300 years of weekly data. Because of item 2 the two horizons imply materially different statistics. & The published figure cannot be reproduced from the archive, and the discrepancy is quantitatively large. \\

    4 & Appendix F, Equations (F1)--(F3) & BF replace the model with an \iid one-dynasty multiplier independent of current wealth. In the BF economy the return depends on belief and the endogenous market price, both coupled with wealth, so the factorization $\varrho(W)p(\eta)$ in (F2) fails. & Equation (F3) is not a tail equation for the BF wealth process. \\

    5 & Appendix F; Equations (5), (6), (33) & Conditional on birth, survival, and the public history a dynasty receives no new idiosyncratic multiplicative shock, and with $\lambda=\sqrt\delta$ beliefs homogenize within a few years of a fifty-year life. Large $N$ averages deaths and newborn priors but not the common signal history. & Both renewal conditions behind (F1) fail, and the Manrubia--Zanette calculation does not justify a large-$N$ cross-sectional Pareto law. \\

    6 & Appendix F, Equations (F4)--(F5); Equation (38) & Both formulas assume small $\sigma^2$. Measured at BF's calibration, $\bar\sigma^2\approx\numSigmaBarSq$, and substitution returns $\mu\approx\numMuFfour$, not the reported $1.4$. Appendix F also imposes $\E[\eta]=0$ and identifies $\bar\sigma^2$ with the variance, whereas Equation (38) computes $\E[\eta^2]$ and Equation (37) gives a generally nonzero conditional mean, contributing a first-order term $\mu\E[\eta]$ that BF do not justify dropping. At BF's calibration $\E[\eta]=\numMeanEta$ exceeds $\delta$ itself, so the omitted term is not negligible. & BF's stated analytical support does not reproduce their own exponent. \\

    7 & Appendix F, Equation (F5) & Under a block-constant formalization with $L\approx\lambda^{-1}$ the tail condition is $(1-\delta)^L\E[(1+\eta)^{\mu L}]=1$, whose second-order solution is \eqref{eq:blockmu}, not (F5). The two agree only in the regime where the additive constants are negligible. & The guarantee $\mu\geq1$ is an artifact of the algebraic form of (F5). \\

    8 & Page 980, footnote 28 & The nondegenerate $\delta\to0$ limit of $1.23$ requires $\delta/\bar\sigma^2\to1$. The measured ratio at BF's own calibration is $\numDeltaOverSigmaSq$. & The small-mortality conclusion is unsupported; the same formula gives a degenerate limit. \\

    9 & Page 967; Equations (19)--(21), (25), and (30) & BF call their Arrow-security allocation rule ``[t]he novel aspect of our approach.'' \citet[p.~565]{Rubinstein1976} had already derived the same state-contingent log-utility rule and its wealth-weighted consensus belief; \citet{DetempleMurthy1994} and \citet{JouiniNapp2007} give intertemporal and complete-market formulations. These antecedents are not cited by BF. & The closed-form decision and pricing rules specialize established results, so the claimed novelty of the market block is unsupported. \\
\end{longtable}
\endgroup

\begingroup
\footnotesize
\setlength{\tabcolsep}{3.5pt}
\begin{longtable}{@{}c L{0.16\textwidth} L{0.43\textwidth} L{0.29\textwidth}@{}}
    \caption{Inventory of additional issues in \citet{BouchaudFarmer2023}.}\label{tab:inventory-additional}\\
    \toprule
    No. & Location & Issue or unsupported step & Consequence \\
    \midrule
    \endfirsthead

    \multicolumn{4}{l}{\tablename\ \thetable\ (continued)}\\
    \toprule
    No. & Location & Issue or unsupported step & Consequence \\
    \midrule
    \endhead

    \midrule
    \multicolumn{4}{r}{Continued on next page}\\
    \endfoot

    \bottomrule
    \endlastfoot

    1 & Replication lines 63, 67--68, 90, 96--98 & The program includes realized survival indicators in the price numerator, divides by $NH$ instead of $N_xH$, and never uses the survivor count it computes, mixing BF's Equation (25) numerator with the Equation (32) denominator. & The code violates BF's finite-population pricing formula and the aggregate wealth identity, though the quantitative effect is negligible. \\

    2 & Replication lines 92--103 & The program draws $s_t$, uses it to update beliefs to $p_{i,t+1}$, and then pays the return on $s_t$ using those updated beliefs, whereas the model prices claims on $s_t$ with $p_{i,t}$. & The code uses post-outcome beliefs to settle claims contingent on the same outcome. \\

    3 & Page 955, Theorem 1 and following paragraph & The stated theorem omits the requirement that the shift preserve the probability measure, and gives $\E(f\mid\mathcal I)$, which can be random, whereas BF say the average converges to the marginal stationary mean. That requires ergodicity. & The statement conflates stationarity, existence of time averages, and ergodicity. \\

    4 & Replication lines 16 and 124--135; p.~950, footnote 2 & The tail interval $7\leq\log W\leq12$ is chosen in advance with no threshold analysis or regular-variation diagnostic, and BF define a Pareto tail as a Pareto fit beyond two standard deviations rather than as asymptotic regular variation. & The fitted slope neither establishes a Pareto tail nor identifies an exponent; alternative tail observations give materially different estimates. \\

    5 & Replication line 8 and footnote 22 & BF report one fixed-seed history and state that unreported experiments suggest limited variability, but neither the article nor the archive provides a Monte Carlo summary. The public signal remains aggregate as $N\to\infty$. & The reported exponent and Gini coefficient do not convey variation across signal histories. \\
\end{longtable}
\endgroup

\end{document}